\documentclass[conference]{IEEEtran}
\IEEEoverridecommandlockouts
\usepackage{cite}
\usepackage{makecell}
\usepackage{multirow, multicol,makecell}
\usepackage{amsmath,amssymb,amsfonts}
\usepackage{algorithmic}
\usepackage{graphicx}
\usepackage{textcomp}
\usepackage{graphicx}
\usepackage{threeparttable,booktabs}
\usepackage{xcolor}
\usepackage{caption}
\usepackage{diagbox}
\usepackage{comment}
\usepackage{romannum}
\usepackage{color}
\usepackage[dvipsnames]{xcolor}

\usepackage[skip=3pt,font=footnotesize]{caption}
\usepackage{subcaption}
\DeclareCaptionTextFormat{uppercase}{\MakeUppercase{#1}}
\usepackage{amsmath,amsfonts}
\usepackage{algorithmic}
\usepackage{algorithm}
\usepackage{array}
\usepackage{textcomp}
\usepackage{stfloats}
\usepackage{url}
\usepackage{verbatim}
\usepackage{graphicx}
\usepackage{cite}
\definecolor{mypink1}{rgb}{0.858, 0.188, 0.478}
\def\BibTeX{{\rm B\kern-.05em{\sc i\kern-.025em b}\kern-.08em
    T\kern-.1667em\lower.7ex\hbox{E}\kern-.125emX}}

\begin{document}

\title{End-to-End Neural Decomposition with Koopman Operators for Time-Series Forecasting\\
}

\author{\IEEEauthorblockN{De-Yan Lu\IEEEauthorrefmark{1}, Xugang Lu\IEEEauthorrefmark{2}, Yu Tsao\IEEEauthorrefmark{3} and Jian-Jiun Ding\IEEEauthorrefmark{1}}\IEEEauthorblockA{\IEEEauthorrefmark{1}Graduate Institute of Communication Engineering, National Taiwan University, Taipei, Taiwan\\
Email: qe59979022@gmail.com, jjding@ntu.edu.tw}
\IEEEauthorblockA{\IEEEauthorrefmark{2}National Institute of Information and Communications Technology, Kyoto, Japan\\
Email: xugang.lu@nict.go.jp}
\IEEEauthorblockA{\IEEEauthorrefmark{3}Research Center for Information Technology Innovation, Academic Sinica, Taipei, Taiwan\\
Email: yu.tsao@citi.sinica.edu.tw}}
 
\maketitle

\begin{abstract}
Koopman theory offers a linear-operator view of nonlinear sequence dynamics by lifting observations into a space where evolution is governed by a linear time-invariant Koopman operator. While the Koopman operator provides a linear representation of nonlinear dynamics, it is generally infinite-dimensional and defined under time-invariant assumptions. To model non-stationary signals with frequency-dependent behavior, a frequency-varying extension is required. In recent years, deep learning has been increasingly employed to exploit its powerful function-approximation ability for learning the Koopman operator. In this study, we propose a novel approach called neural decomposition Koopman (NDKoop), an end-to-end architecture that integrates a learnable signal decomposition module with both frequency-independent and frequency-dependent Koopman-based networks for sequence forecasting. To the best of our knowledge, this is the first work to jointly realize end-to-end Koopman modeling and signal decomposition within a unified neural framework. We demonstrate that decomposing a signal into a frequency-independent trend component and a frequency-dependent periodic component, each governed by a corresponding Koopman operator, improves prediction accuracy when perfect linearization is unattainable. Numerical experiments across several forecasting benchmarks indicate that the proposed NDKoop provides strong performance.

\end{abstract}

\section{Introduction}
The prediction and analysis of nonlinear dynamical systems remain challenging problems in many scientific and engineering domains. Koopman theory \cite{C1,C2} provides a principled framework for representing nonlinear dynamics as linear evolution in a space of infinite-dimensional measurement functions. Although Koopman theory provides an elegant theoretical framework, the associated infinite-dimensional function space is infeasible to construct in practice, thereby requiring finite-dimensional approximations of the Koopman operator. While Dynamic Mode Decomposition (DMD) \cite{C3,C4,C5} provides a computationally efficient approximation of the Koopman operator, it exhibits notable drawbacks. In particular, DMD is sensitive to measurement noise, struggles to accurately capture strongly nonlinear or transient dynamics, and is constrained by its reliance on a fixed set of linear observables.

Recently, deep learning has demonstrated remarkable success in sequence forecasting, including MLP-based approaches such as DLinear \cite{C6,C7}. However, these methods often lack explicit temporal structure modeling and exhibit limited capability in capturing long-range dependencies and nonstationary dynamics. Although computationally efficient, their inherent linear assumptions restrict their ability to model nonlinear and frequency-varying behaviors effectively. Transformer-based models, such as PatchTST \cite{C8}, have further advanced long-term forecasting performance by improving the efficiency of the attention mechanism. Nevertheless, transformers and other deep neural networks may still suffer from limited generalization under distribution shifts and increased computational cost when scaling to large collections of time series. In contrast, Koopman-based approaches explicitly characterize the underlying system dynamics and provide interpretable state transitions under both time-invariant and time-varying settings, including representative methods such as Koopa \cite{C9}. However, Koopa relies on manually designed Fourier filtering with predefined frequency thresholds, which limits adaptability across datasets and reduces robustness under noisy or highly nonstationary conditions.

In this study, we propose a novel Neural Decomposition Koopman (NDKoop) framework that integrates neural networks (NNs), Koopman operator theory, and signal decomposition to achieve interpretable latent linear dynamics. NDKoop shows strong potential in biomedical signal analysis, industrial monitoring, energy forecasting, financial prediction, and other complex dynamical systems by effectively modeling nonlinear and nonstationary temporal dynamics with high interpretability and computational efficiency. To our knowledge, no prior work has jointly realized signal decomposition and Koopman-based modeling within a fully end-to-end neural architecture. Based on the proposed framework, we develop two Koopman models: a frequency-independent Koopman network and a frequency-dependent Koopman network to capture global and frequency-specific dynamics, respectively. Trend-like components are modeled using a shared time-invariant Koopman operator, while oscillatory components employ frequency-specific operators to represent distinct periodic behaviors. By separating these dynamics, NDKoop more effectively captures global trends and temporal patterns, improving forecasting accuracy, robustness, and interpretability for nonstationary time series.

\section{Proposed Neural Decomposition Koopman (NDKoop) Framework}
This section formulates the problem, analyzes frequency-independent and frequency-dependent Koopman operators, and presents the proposed NDKoop framework.

\subsection{Problem Statement}
Consider a discrete-time dynamical system that generates a trajectory $\{x_k\}_{k=1}^L$. The system states are latent and cannot be directly observed. Instead, we have access to measurements $y_k=g(x_k)$, obtained through an unknown observation function $g$. 

Given a dataset comprising multiple sequences of length $L+T$, the forecasting task is to estimate the future measurements $y_{L+1},y_{L+2},...,y_{L+T}$ based solely on the observed prefix $y_1,y_2,...,y_L$.

\subsection{Fundamentals of Frequency-Independent and Frequency-Dependent Koopman Operators}
The Koopman operator is linear and time-independent. For simple signals, such as linear trends $x(t)$, the measurement function (encoder $\phi$) lifts the one-dimensional signal into a high-dimensional latent space (e.g., two dimensions) where the dynamics are governed by a time-invariant, frequency-independent Koopman operator $\mathbf{K}$. The operator propagates the latent state forward in time, and a decoder $\phi^{-1}$ maps the predicted latent state back to the original signal space. An overview is shown in Fig. 1(a). The linear trend $x(t)$, latent feature $\mathbf{z}(t)$ and time-frequency-independent (frequency-independent) Koopman operator $\mathbf{K}$ can be expressed as follows:
\begin{equation}
x(t)=at+b, \quad
\mathbf{z}(t) =
\begin{bmatrix}
x(t) \\
1 \\
\end{bmatrix},\quad  
\mathbf{K} = 
\begin{bmatrix}
1 & a \\
0 & 1
\end{bmatrix},\quad a,b\in\mathbb{R},  \label{eq}
\end{equation}
While the input signal $x(t)$ exhibits sinusoidal behavior, the resulting Koopman operator becomes explicitly frequency-dependent. An overview of this process is illustrated in Fig. 1(b). The input signal $x(t)$, latent feature $\mathbf{z}(t)$ and time-independent but frequency-dependent (frequency-dependent) Koopman operator $\mathbf{K}(\omega)$ can be expressed as follows:
\begin{equation}
x(t)=\cos{(\omega t)}, \quad
\mathbf{z}(t) =
\begin{bmatrix}
\cos{(\omega t)} \\
\sin{(\omega t)} \\
\end{bmatrix},\label{eq} 
\end{equation}

\begin{equation}
\mathbf{K}(\omega) = 
\begin{bmatrix}
\cos{(\omega t)} & -\sin{(\omega t)} \\
\sin{(\omega t)} & \cos{(\omega t)}
\end{bmatrix},\quad \omega\in\mathbb{R},  \label{eq}
\end{equation}
where $\omega$ is the angular frequency of the signal.

\begin{figure}[!t]
\captionsetup{labelfont=normalfont,labelsep=period}
\centering
\begin{subfigure}{\linewidth}
  \centering
  \includegraphics[width=9.0cm]{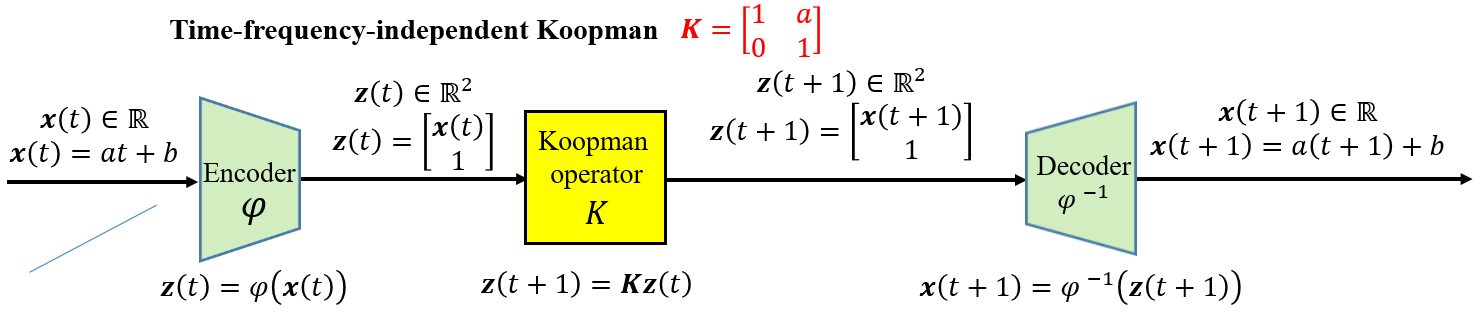}
  \caption{Frequency-independent Koopman for a line trend}
\end{subfigure}

\vspace{0.5em}

\begin{subfigure}{\linewidth}
  \centering
  \includegraphics[width=9.0cm]{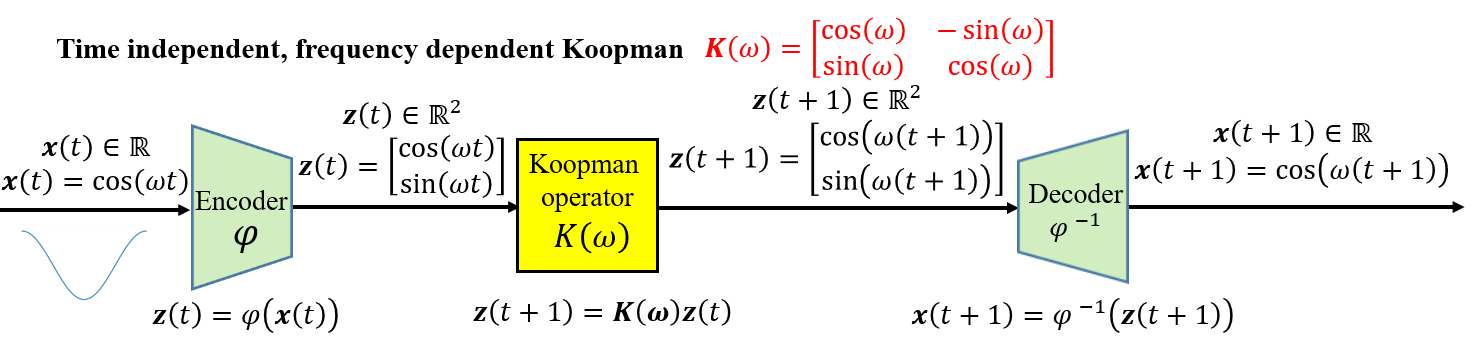}
  \caption{Frequency-dependent Koopman for a sinusoid}
\end{subfigure}

\caption{Frequency-independent and frequency-dependent Koopman}
\end{figure}

\begin{figure*}[!t]
    \captionsetup{labelfont=normalfont, justification=centering, labelsep=period}
    \centering 
    \centerline{\includegraphics[width=15.3cm]{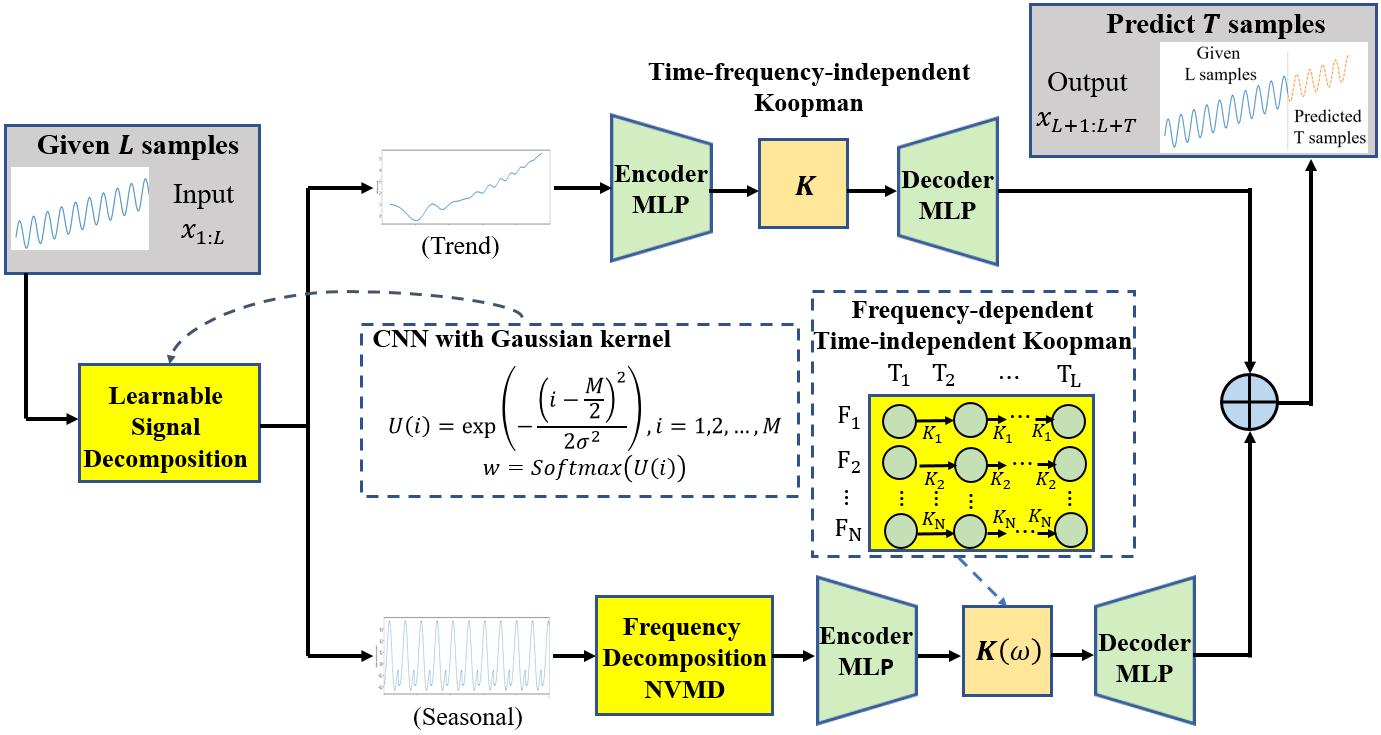}}
    \caption{The proposed NDKoop framework.}
    \label{fig:NDKOOP framework}
\end{figure*}

\subsection{Architecture of the NDKoop System}
The proposed NDKoop framework, illustrated in Fig. 2, consists of a learnable signal decomposition module, a frequency decomposition stage based on neural variational mode decomposition (NVMD), and both frequency-independent and frequency-dependent Koopman NNs. In the following, we describe each component in detail and explain its role within the overall architecture.

\subsubsection{Learnable Signal Decomposition Module}
To implement the learnable convolutional decomposition, we first define a one-dimensional convolutional kernel \cite{C10} with stride $S=2$ and kernel size $M=20$, selected empirically. The kernel weights are initialized using a Gaussian distribution \cite{C11}. Specifically, we define $U \in \mathbb{R}^{M}$ with entries
\begin{equation}
U[i] = \exp\!\left(-\frac{(i - M/2)^2}{2\sigma^2}\right),
\quad i = 1, \ldots, M,  \label{eq}
\end{equation}
where $\sigma \in\mathbb{R}$ is a hyperparameter set to 1. The convolutional kernel weights are then obtained by applying a softmax operation along the kernel dimension, i.e.,
\begin{equation}
w=Softmax(U),  \label{eq}
\end{equation}
Compared with a moving average filter \cite{C12,C13}, a Gaussian kernel provides smoother weighting with greater emphasis on central samples, leading to improved locality preservation and reduced boundary artifacts. As a result, the learnable signal decomposition (LSD) yields the long-term trend component $x_{trend}(t)$, and the remaining residual represents the periodic seasonal component $x_{season}(t)$, given by

\begin{equation}
    x_{trend}(t) = LSD(x(t)), \quad x_{season}(t)=x(t)-x_{trend}(t), \label{eq}
\end{equation}
where $x(t)$ is the input signal.
\subsubsection{Frequency Decomposition NVMD Module}
The conventional VMD \cite{C14} decomposes a signal $x(t)$ into $N$ sub-signals $\{u_i(t)\}_{i=1}^N$, each of which is narrowband and concentrated around a corresponding center frequency, exhibiting quasi-sinusoidal behavior.
\begin{equation}
x(t)=\sum_{i=1} ^{N} u_i(t), \quad u_i(t) = A_i(t)\cos\!\big(\phi_i(t)\big),  \label{eq}
\end{equation}
where $A_i(t)$ is the amplitude of the $i^{th}$ subband signal, and $\phi_i(t)$ is the phase of the $i^{th}$ subband signal. Since periodic signals can be represented as a superposition of fundamental frequency components, VMD is well suited for performing frequency-based signal decomposition.

Despite its effectiveness, conventional VMD entails high computational cost due to its iterative optimization procedure. To address this limitation, we employ a neural network–based surrogate, termed NVMD \cite{C15}, which approximates the behavior of traditional VMD while substantially reducing computational complexity.

\begin{figure}[!t]
\captionsetup{labelfont=normalfont,labelsep=period}
\centering
\begin{subfigure}{\linewidth}
  \centering
  \includegraphics[width=9.0cm]{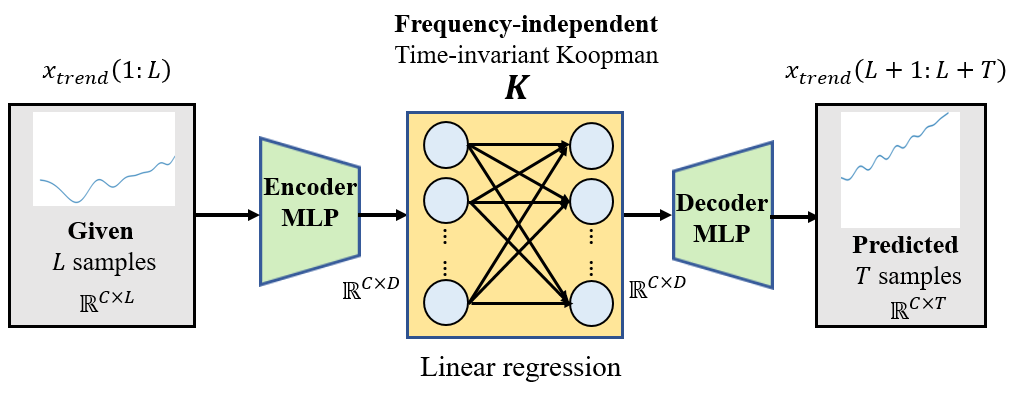}
  \caption{Architecture of frequency-independent Koopman operator}
\end{subfigure}

\vspace{0.5em}

\begin{subfigure}{\linewidth}
  \centering
  \includegraphics[width=9.2cm]{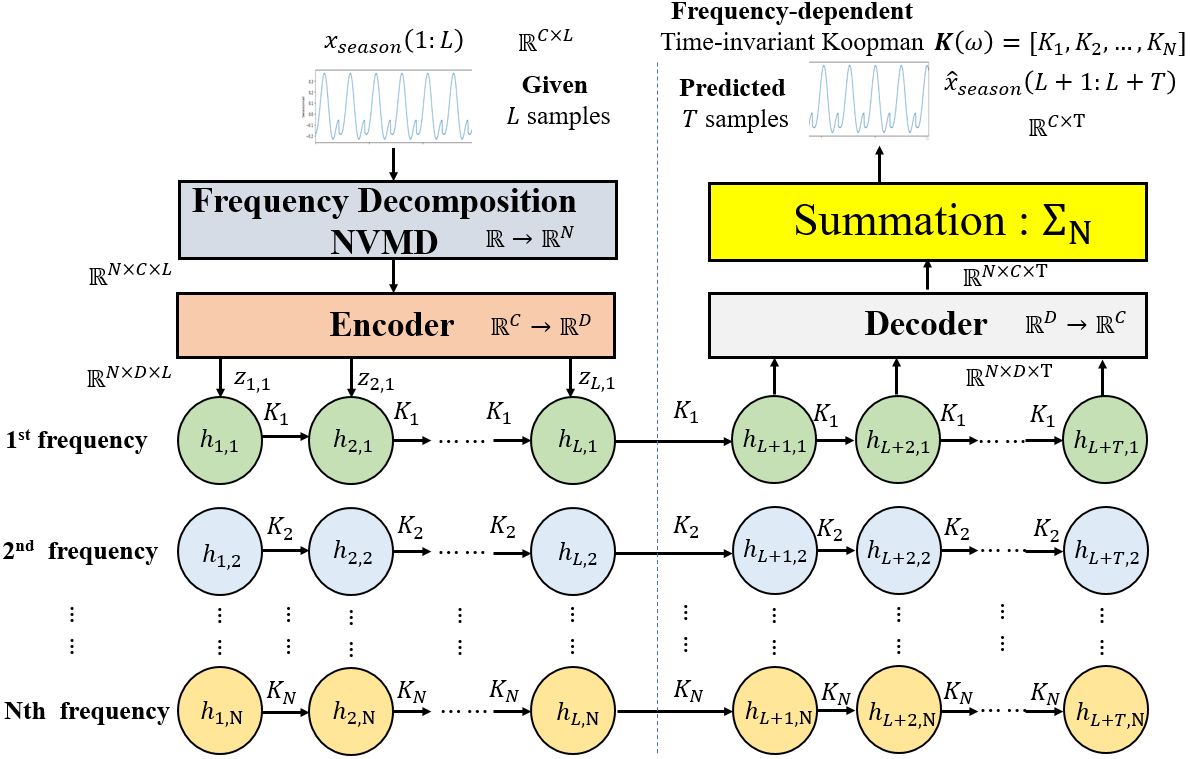}
  \caption{Architecture of frequency-dependent Koopman operator}
\end{subfigure}

\caption{Frequency-independent and dependent Koopman NNs.}
\end{figure}

\subsubsection{Frequency-Independent and Frequency-Dependent Koopman NNs}
This section introduces the frequency-independent Koopman NN for modeling non-oscillatory and slowly varying time-series components, such as trends. These dynamics are captured by a frequency-independent linear operator in a lifted latent space, enabling structured and interpretable long-horizon forecasting. We also present a frequency-dependent Koopman framework for modeling oscillatory and multi-frequency dynamics, such as seasonality. By combining NVMD with frequency-specific Koopman operators, the proposed approach preserves linear latent evolution while effectively capturing heterogeneous temporal behaviors.

Fig. 3(a) llustrates the architecture of the proposed frequency-independent Koopman network. The input trend sequence $\{x_{trend,k}\}_{k=1}^L \in \mathbb{R}^{C \times L}$, where $C$ denotes the number of channels and $L$ the sequence length, is first mapped into a latent representation through an MLP-based encoder, which lifts the signal from the original observation space $\mathbb{R}^{C}$ to a latent space $\mathbb{R}^{D}$. In this latent space, the temporal evolution is governed by a time-invariant Koopman operator $\mathbf{K}$, implemented as a linear transformation. This operator captures the global linear dynamics of the trend component and is shared across all time steps. During the prediction phase, the Koopman operator recursively propagates the latent state forward in time to generate latent predictions for the next $T$ steps. These predicted latent states are then mapped back to the original signal space through an MLP-based decoder, producing the forecasted trend sequence $\{\hat{x}_{trend,k}\}_{k=L+1}^{L+T} \in \mathbb{R}^{C \times T}$.

Fig. 3(b) illustrates the overall architecture of the proposed frequency-dependent Koopman network. The input seasonal sequence $\{x_{season,k}\}_{k=1}^L \in \mathbb{R}^{C \times L}$ is first processed by NVMD, which decomposes the signal into $N$ frequency components. This operation produces a tensor of decomposed signals with dimension $\mathbb{R}^{N \times C \times L}$, where each component corresponds to a distinct oscillatory mode. Each frequency component is independently mapped into a shared latent space through a learnable encoder, which transforms the input from $\mathbb{R}^{C}$ to $\mathbb{R}^{D}$. In the latent space, the temporal evolution of each frequency mode is governed by a time-invariant but frequency-dependent Koopman operator. Specifically, the latent state update for the $n^{th}$ frequency component is given by
\begin{equation}
    \mathbf{h}_{k+1,n} = \mathbf{K}_n \mathbf{h}_{k,n} + \mathbf{z}_{k,n}, \label{eq}
\end{equation}
\begin{equation}
    n=1,2,...,N, \quad k=1,2,...,L, \label{eq}
\end{equation}
where $\mathbf{K}_n$ is the Koopman operator associated with the $n^{th}$ frequency, and $\mathbf{z}_{k,n}$ is the encoded input. 

During the prediction phase, the learned Koopman operators $\{\mathbf{K}_n\}_{n=1}^N$ propagate the latent states forward in time to generate latent predictions over the next $T$ steps, given by
\begin{equation}
    \mathbf{h}_{k+1,n} = \mathbf{K}_n \mathbf{h}_{k,n}, \label{eq}
\end{equation}
\begin{equation}
    n=1,2,...,N, \quad k=L+1,...,L+T-1, \label{eq}
\end{equation}
These predicted latent trajectories are then mapped back to the original signal space through a shared decoder, producing frequency-wise forecasts of dimension $\mathbb{R}^{N \times C \times T}$. Finally, the predicted components across all frequencies are aggregated via a summation operation to obtain the final forecast seasonal sequence $\{\hat{x}_{season,k}\}_{k=L+1}^{L+T}\in \mathbb{R}^{C \times T}$. Therefore, the final predicted sequence  representation $\{\hat{x}_{k}\}_{k=L+1}^{L+T}\in \mathbb{R}^{C \times T}$ with the next $T$ time steps is given by
\begin{equation}
    \{\hat{x}_{k}\}_{k=L+1}^{L+T}=\{\hat{x}_{trend,k}\}_{k=L+1}^{L+T}+\{\hat{x}_{season,k}\}_{k=L+1}^{L+T}, \label{eq}
\end{equation}

\subsection{Forecasting Objective}
In NDKoop, the parameters of the learnable signal decomposition, encoder, decoder, NVMD, the frequency-independent Koopman operator, and the frequency-dependent Koopman operator are denoted as $\theta_{\mathrm{LD}}$, $\theta_{\mathrm{E}}$, $\theta_{\mathrm{D}}$, $\theta_{\mathrm{NVMD}}$, \(\mathbf{K}\), and \(\mathbf{K}(\omega)\), respectively. Model parameters are optimized by minimizing the mean squared error (MSE) between the predicted sequence \(\hat{y}\) and the ground-truth sequence \(y\), formulated as
\begin{equation}
\arg\min_{\theta_{\mathrm{LD}},\theta_{\mathrm{E}},\theta_{\mathrm{D}},\theta_{\mathrm{NVMD}},\mathbf{K},\mathbf{K}(\omega)}
\mathcal{L}_{\mathrm{MSE}}(\hat{y}, y),
\label{eq:forecasting_objective}
\end{equation}

\begin{table}[!t]
\centering
\caption{Multivariate forecasting results under different forecast horizons $T$. The lookback length is set to $L = 2T$.}
\small
\renewcommand{\arraystretch}{1.1} 
\setlength{\tabcolsep}{3pt}      
\resizebox{\linewidth}{!}{
\begin{tabular}{lc|cc cc cc cc}
\toprule
\multicolumn{2}{c}{Models} & \multicolumn{2}{c}{\textbf{NDKoop}} & \multicolumn{2}{c}{Koopa \cite{C9}} & \multicolumn{2}{c}{PatchTST \cite{C8}} & \multicolumn{2}{c}{DLinear \cite{C6,C7}}  \\
\cmidrule(lr){1-2} \cmidrule(lr){3-4} \cmidrule(lr){5-6} \cmidrule(lr){7-8} \cmidrule(lr){9-10} 
\multicolumn{2}{c}{Metric} & MSE & MAE & MSE & MAE & MSE & MAE & MSE & MAE  \\
\midrule
\multirow{4}{*}{\rotatebox{90}{ECL}} 
& 48  &\textbf{0.121}  &\textbf{0.211}  & \underline{0.130} & \underline{0.234} & 0.147 & 0.246  & 0.158 & 0.241  \\
& 96  &\textbf{0.129} &\textbf{0.219}   &  \underline{0.136} & \underline{0.236} & 0.143 & 0.241  & 0.153 & 0.245  \\
& 144 &\textbf{0.137}  &\underline{0.242}  &0.149 & 0.247 &  \underline{0.145} & \textbf{0.241}  & 0.152 & 0.245  \\
& 192 &\textbf{0.146}  &\underline{0.244}  & 0.156 & 0.254 & \underline{0.147} & \textbf{0.240}  & 0.153 & 0.246  \\
\midrule
\multirow{4}{*}{\rotatebox{90}{ETTh2}} 
& 48 &\textbf{0.221}  &\textbf{0.293}  & 0.226 & 0.300 & \underline{0.223} & \underline{0.297}  & 0.226 & 0.305  \\
& 96 &\textbf{0.289}  &\textbf{0.340}  & 0.297 & \underline{0.349} & 0.300 & 0.353  & \underline{0.294} & 0.351  \\
& 144 &\textbf{0.321}  &\textbf{0.369}  & \underline{0.333} & \underline{0.381} & 0.346 & 0.390  & 0.354 & 0.397  \\
& 192 &\textbf{0.345}  &\textbf{0.385}  & \underline{0.356} & \underline{0.393} & 0.383 & 0.406  & 0.385 & 0.418  \\
\midrule
\multirow{4}{*}{\rotatebox{90}{ILI}} 
& 24 &\textbf{1.526}  &\textbf{0.743}  & \underline{1.621} & \underline{0.800} & 2.063 & 0.881  & 2.624 & 1.118  \\
& 36  &\textbf{1.413}  &\textbf{0.708}  & \underline{1.803} & \underline{0.855} & 2.178 & 0.943  & 2.693 & 1.156  \\
& 48  &\textbf{1.494}  &\textbf{0.782}  & \underline{1.768} & 0.903 & 1.916 & \underline{0.896}  & 2.852 & 1.229  \\
& 60  &\underline{1.893}  &0.945  & \textbf{1.743} & \textbf{0.891} & 1.981 & \underline{0.917}  & 2.554 & 1.144  \\
\midrule
\multirow{4}{*}{\rotatebox{90}{ECG}} 
& 48  &\textbf{0.245}  &\textbf{0.301}  &\underline{0.277}  &\underline{0.322}  &0.313  & 0.375  &0.344  &0.414   \\
& 96  &\textbf{0.271}  &\textbf{0.324}  &\underline{0.291}  &\underline{0.338}  &0.357  &0.403  &0.381  &0.435   \\
& 144 &\underline{0.322}  &\textbf{0.354}  &\textbf{0.320}  &\underline{0.361}  &0.389  &0.434   &0.406  &0.462   \\
& 192 &\underline{0.364}  &\textbf{0.391}  &\textbf{0.349}  &\underline{0.394}  &0.410  &0.479   &0.448  &0.547   \\
\midrule

\multicolumn{2}{l}{1\textsuperscript{st} Count} & 13 & 13 & 3 & 1 & 0 & 2 & 0 & 0 \\
\bottomrule
\end{tabular}
}
\end{table}

\begin{figure}[!t]
\captionsetup{labelfont=normalfont, labelsep=period}
\centering 
\centerline{\includegraphics[width=8.5cm]{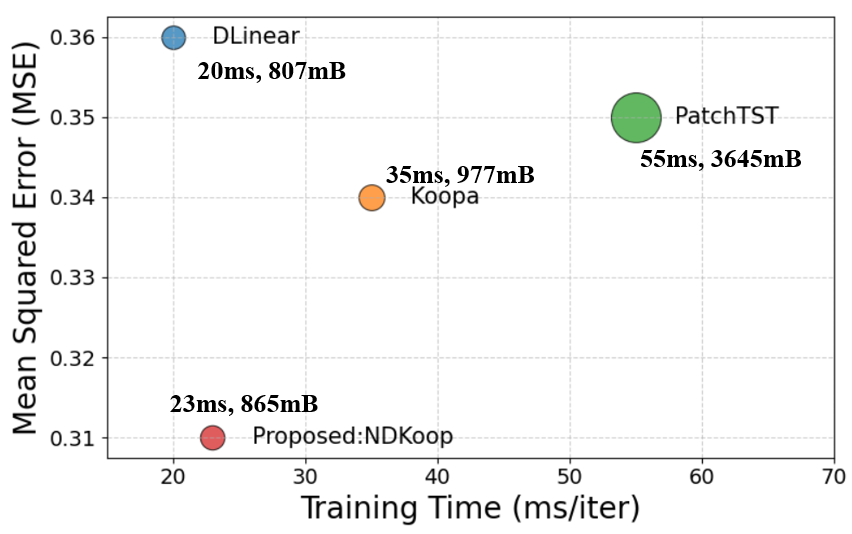}}
\caption{Model efficiency comparison on error and training time per epoch. Memory consumption is proportional to circle radius.}
\label{fig}
\end{figure}

\section{Experiments}
\subsection{Experimental Setup}
We evaluated NDKoop on four real-world benchmarks widely adopted in forecasting tasks, including ECL, ETTh2, ILI, and ECG (MIT-BIH) \cite{C16}, following the same data preprocessing procedures and train–test split settings as Koopa. Instead of using a fixed lookback window, for each forecast horizon $T$, we set the lookback length to $L=2T$. This design reflects real-world scenarios where longer histories are available and allows deep models to exploit additional observations as the forecasting horizon increases. 

NDKoop was compared with a broad set of state-of-the-art baselines, including Transformer-based (PatchTST), MLP-based (DLinear), and Koopman-based (Koopa) methods. All baselines were reproduced using their official implementations or descriptions in the original papers. Each experiment was conducted with three different random seeds, and the average test MSE and mean absolute error (MAE) are reported.

\subsection{Performance Evaluation Criteria}
In this study, the average MSE and MAE are used as quantitative  performance estimators on test data. A smaller MSE and MAE for the forecasting task indicate better predictive accuracy. The average MSE and MAE can be calculated from:
\begin{equation}
\mathrm{MSE} = \frac{1}{T} \sum_{i=1}^{T} \left( y_i - \hat{y}_i \right)^2, \quad \mathrm{MAE} = \frac{1}{T} \sum_{i=1}^{T} \left| y_i - \hat{y}_i \right|,
\end{equation}
where $y_i$ is the ground truth sequence, and $\hat{y}_i$ is the corresponding predicted sequence. MSE and MAE are widely adopted and fundamental evaluation metrics for forecasting tasks. Nevertheless, incorporating additional evaluation criteria, such as correlation-based, similarity-based, and signal-quality-related metrics, could provide a more comprehensive and convincing assessment of model performance in future studies.

\subsection{Experimental Results}
Table I summarizes the multivariate forecasting results under different prediction horizons on the ECL, ETTh2, ILI, and ECG datasets, evaluated using MSE and MAE (lower is better). The lookback length is fixed to $L=2T$. The ILI dataset is sampled at a weekly frequency and exhibits pronounced nonstationarity, resulting in a comparatively shorter effective temporal scale. Accordingly, the forecasting horizons are set to $T \in \{24,36,48,60\}$. In contrast, for the other datasets, we adopt $T \in \{48,96,144,192\}$. The variation in forecasting horizons across datasets reflects differences in sampling frequency and established benchmark protocols. Specifically, shorter horizons are employed for ILI in accordance with prior epidemiological forecasting studies, whereas longer horizons are used for ECL, ETTh2, and ECG to ensure consistency with standard long-term time-series forecasting benchmarks.

Overall, NDKoop consistently achieves the best performance across most datasets and horizons, particularly in terms of MSE. On the ECL and ETTh2 datasets, NDKoop outperforms all baselines for all forecasting horizons, with performance gaps increasing as the horizon grows. For the ILI dataset, which exhibits strong nonstationarity, NDKoop shows substantial advantages over competing methods, especially in long-term forecasting. On the ECG dataset, NDKoop attains the lowest errors for most horizons, demonstrating its effectiveness in modeling complex biomedical time-series dynamics. While PatchTST and Koopman-based baselines perform competitively for short horizons, their performance degrades more rapidly for longer horizons. PatchTST and DLinear exhibit weaker performance because they primarily rely on attention-based or linear trend modeling, respectively, without explicitly enforcing latent linear dynamics, making them less effective for long-term and highly nonstationary time-series forecasting. In contrast, NDKoop maintains stable accuracy across extended prediction lengths, indicating superior robustness and generalization. Most importantly, NDKoop employs a fully learnable end-to-end architecture and explicitly models distinct Koopman structures across different frequency components, providing a clear advantage over Koopa.

Fig. 4 illustrates the trade-off between forecasting accuracy and computational efficiency. NDKoop achieves the lowest MSE and shortest training time, while requiring less memory than Transformer-based baselines. Although PatchTST attains competitive accuracy, it incurs higher computational and memory costs due to quadratic attention complexity. Koopa also exhibits higher computational cost than NDKoop because it employs multiple neural operators and frequency decomposition stages, increasing model complexity and training overhead. In contrast, DLinear is efficient but delivers inferior accuracy due to limited modeling capacity. Overall, NDKoop provides the most favorable balance among accuracy, efficiency, and memory usage.

\section{Conclusion}
This study proposes a novel learning-based method, NDKoop, for time-series prediction. NDKoop maps signals into frequency-independent and frequency-dependent linear latent spaces, enabling effective identification of Koopman operators and measurement functions while reducing overall training complexity. As a result, NDKoop achieves superior efficiency and performance compared with existing MLP-based, Transformer-based, and Koopman-based methods in signal analysis and time-series forecasting. Therefore, NDKoop provides a highly interpretable architecture by integrating signal decomposition with Koopman theory, substantially enhancing the interpretability of neural networks.


\vspace{12pt}
\end{document}